\documentclass[letterpaper, 10 pt, conference]{ieeeconf}
\IEEEoverridecommandlockouts
\usepackage{graphics} % for pdf, bitmapped graphics files
\usepackage{epsfig} % for postscript graphics files
\usepackage{mathptmx} % assumes new font selection scheme installed
\usepackage{amsmath} % assumes amsmath package installed
\usepackage{amssymb}  % assumes amsmath package installed
\usepackage{multirow}
\usepackage{bbding}
\usepackage{graphicx}
\usepackage{subcaption}
\usepackage{flushend}
\usepackage{hyperref}
\usepackage{xcolor}
\usepackage{cite}
\usepackage{jabbrv}

\hypersetup{
    colorlinks,
    linkcolor={red!50!black},
    citecolor={blue!50!black},
    urlcolor={blue!80!black}
}

\title{\LARGE \bf
In the driver's mind: modeling the dynamics of human overtaking decisions in interactions with oncoming automated vehicles
}

\author{Samir H.A. Mohammad$^{1,*}$, Haneen Farah$^{2}$ and Arkady Zgonnikov$^{1}$% <-this % stops a space
\thanks{$^{1}$Samir H.A. Mohammad* and Arkady Zgonnikov are with the Department of Cognitive Robotics, Faculty of Mechanical Engineering, Delft University of Technology, Mekelweg 2, Delft, 2628CD, the Netherlands.
        {\tt\small s.h.a.mohammad@tudelft.nl, a.zgonnikov@tudelft.nl}
        }%
\thanks{$^{2}$Haneen Farah is with the Department of Transport and Planning, Faculty of Civil Engineering and Geosciences, Delft University of Technology, Stevinweg 1, Delft, 2628CN, the Netherlands.
        {\tt\small h.farah@tudelft.nl}
        }%
\thanks{*Corresponding author.}% E-mail address: s.h.a.mohammad@tudelft.nl (S.H.A. Mohammad)}
\thanks{All the data and code produced in this study, as well as online supplementary information are available at \href{https://osf.io/p2wme}{https://osf.io/p2wme}.}%
}

\begin{document}

\maketitle
\thispagestyle{empty}
\pagestyle{empty}
% \renewcommand*{\bibfont}{\normalfont\footnotesize}

%%%%%%%%%%%%%%%%%%%%%%%%%%%%%%%%%%%%%%%%%%%%%%%%%%%%%%%%%%%%%%%%%%%%%%%%%%%%%%%%
\begin{abstract}
Understanding human behavior in overtaking scenarios is crucial for enhancing road safety in mixed traffic with automated vehicles (AVs). Computational models of behavior play a pivotal role in advancing this understanding, as they can provide insight into human behavior generalizing beyond empirical studies. However, existing studies and models of human overtaking behavior have mostly focused on scenarios with simplistic, constant-speed dynamics of oncoming vehicles, disregarding the potential of AVs to proactively influence the decision-making process of the human drivers via implicit communication. Furthermore, despite numerous studies in other scenarios, so far it remained unknown whether overtaking decisions of human drivers are affected by whether they are interacting with an AV or a human-driven vehicle (HDV). To address these gaps, we conducted a ``reverse Wizard-of-Oz'' driving simulator experiment with 30 participants who repeatedly interacted with oncoming AVs and HDVs, measuring the drivers' gap acceptance decisions and response times. The oncoming vehicles featured time-varying dynamics designed to influence the overtaking decisions of the participants by briefly decelerating and then recovering to their initial speed. We found that participants did not alter their overtaking behavior when interacting with oncoming AVs compared to HDVs. Furthermore, we did not find any evidence of brief decelerations of the oncoming vehicle affecting the decisions or response times of the participants. Cognitive modeling of the obtained data revealed that a generalized drift-diffusion model with dynamic drift rate and velocity-dependent decision bias best explained the gap acceptance outcomes and response times observed in the experiment. Overall, our findings highlight the potential of cognitive models for further advancing the ongoing development of safer interactions between human drivers and AVs during overtaking maneuvers.
\end{abstract}

%%%%%%%%%%%%%%%%%%%%%%%%%%%%%%%%%%%%%%%%%%%%%%%%%%%%%%%%%%%%%%%%%%%%%%%%%%%%%%%%
\section{Introduction}
While driving automation offers promise for improving traffic safety \cite{MILAKIS2017324}, successfully managing interactions between automated vehicles (AVs) and human drivers in mixed traffic scenarios remains a substantial challenge \cite{Schieben2019, halting2023}. Overtaking maneuvers on two-lane rural roads, in particular, pose significant risks of head-on collisions at high speeds. Human drivers' inconsistent judgments of available gaps \cite{Gray2005, Lerner2000DriverMO, judgments} underscore the need for a comprehensive understanding of human overtaking behavior to enhance road safety.

Advanced overtaking behavior models can enhance our understanding of human gap acceptance, consequently playing a pivotal role in enhancing road safety and the development of AV technology. These models can simulate realistic human behavior in overtaking scenarios, improving the accuracy of testing and validation of AVs \cite{yu2013modeling}. Furthermore, AVs can utilize such models to predict gap acceptance in real time and anticipate overtaking maneuvers by human drivers, contributing to overall safety \cite{Sadigh2016PlanningFA, schumann2023using}. 

While existing studies contribute to our understanding of overtaking behavior, they predominantly focus on instantaneous decision-making processes (e.g., \cite{Farah2010, SEVENSTER2023329, Gray2005, Llorca2013, Hegeman2005, Polus2000EvaluationOT, farah2016drivers, farah2009passing, Stefansson2020, Vlahogianni2013}), overlooking the dynamic aspects of overtaking interactions. This is in contrast with the dynamic nature of traffic interactions more generally: these interactions evolve over time~\cite{markkula_defining_2020}, influenced by factors such as relative speeds, distances between vehicles, and the behavior of other road users (e.g., negotiations in bottleneck scenarios~\cite{Rettenmaier2020, miller2022implicit} or highway merging~\cite{siebinga_human_2024}). Therefore, models of human overtaking behavior that neglect these dynamic aspects are limited in their applications for human-AV interactions. 

Recent research has started addressing the dynamic aspect of traffic interactions by modeling the dynamics of the decision-making processes across various traffic situations, including pedestrian crossing \cite{Pekkanen2022}, unprotected left turns \cite{ZgonnikovAbbinkMarkkula, nudge, bontje_are_2024}, and overtaking maneuvers \cite{mohammad2023cognitive} using \textit{cognitive process models}, in particular, drift-diffusion models (DDMs). These models assume that drivers integrate visual cues (such as distance and time-to-arrival to oncoming vehicles) over time until sufficient evidence is accumulated. DDMs have effectively incorporated dynamic aspects of interactions, such as an AV signaling yielding intent through deceleration or external human-machine interface signals~\cite{Pekkanen2022, markkula2023explaining}. Furthermore, DDMs have demonstrated potential in describing how time-varying dynamics of oncoming AVs influence gap acceptance decisions and response times \cite{nudge}.

Despite their success in multiple traffic interactions scenarios, DDMs have yet to prove their worth in the context of overtaking interactions. The studies of overtaking have only recently started incorporating advanced measures of human decisions such as response times~\cite{SEVENSTER2023329}; these measures have also informed first attempts to model overtaking using drift-diffusion models~\cite{mohammad2023cognitive}. However, these initial efforts have been limited by their focus on human gap acceptance decisions in response to an oncoming vehicle with trivial, constant-acceleration dynamics. Consequently, human overtaking decisions have yet to be systematically investigated and modeled in the context of dynamic interactions with time-varying kinematics.

Furthermore, it is currently unknown whether overtaking decisions of human drivers are affected by whether they interact with an AV or a human-driven vehicle (HDV). In the context of other traffic interactions, recent studies have presented mixed results on the influence of vehicle type on gap acceptance. Soni et al. \cite{Soni2022} and Trende et al. \cite{trende2019investigation} reported that drivers were willing to accept smaller gaps when interacting with AVs compared to HDVs. However, these studies influenced their participants' perceptions of AVs before the experiment by providing information on expected AV behavior, potentially impacting their results. In contrast, studies that refrained from doing so (e.g., \cite{Reddy2022,velasco2019studying,palmeiro2018interaction}) did not find differences between AVs and HDVs in terms of their effect on gap acceptance. These mixed findings underscore the need for a comprehensive investigation into the potential influence of oncoming vehicle type (AV vs HDV) on human overtaking behavior.

This study aimed to address the above research gaps through cognitive process modeling of data obtained in a driving simulator experiment involving interactions with oncoming AVs and HDVs with time-varying dynamics. The oncoming AV was preprogrammed to perform one of the three behaviors: 1) maintaining a constant speed; 2) decelerating with 2.5 $m/s^2$ for 2 $s$ soon after appearing in participants' field of view and then accelerating back to its original speed (```weak nudge"'); 3) ```strong nudge"' --- similar to the weak nudge but with deceleration magnitude of 5 $m/s^2$. 

To investigate the effect of interacting with AV compared to HDV, we manipulated participants' belief of the type of oncoming vehicle they interacted with. To this end, we used a ``reverse Wizard-of-Oz'' setup: the participants were made to believe that in HDV trials they would interact with the experimenter, while the actual oncoming vehicle was still AV with the same pre-programmed behaviors. 

We hypothesized that a) participants' overtaking behavior (as measured by gap acceptance probability and response time) would remain the same when interacting with AVs compared to HDVs, and b) participants' gap acceptance likelihood in response to deceleration nudges of the oncoming vehicle would be higher, compared to the constant-speed condition. Finally, we fitted four versions of a drift-diffusion model to the collected data to investigate the cognitive mechanisms underlying participants' decisions in dynamic overtaking interactions. 

\section{Methods} \label{methods}
\subsection{Participants}
Approval for this study was granted by the Human Research Ethics Committee of Delft University of Technology. Our participant pool consisted of 30 individuals (15 male, 15 female), with an age range from 18 to 35 years (mean: 24.2, SD: 3.1). On average, participants held a driver's license for 5.6 years, with a range from 0.3 to 17 years (SD = 3.7). Participants' self-reported familiarity with automated vehicles averaged $2.4\pm1.3$ on a 5-point Likert scale. Their self-reported perceived safety of automated vehicles averaged $2.9\pm0.8$. In return for their participation, each participant received a €20 gift voucher.

\begin{figure}[t]
    \centering
    \includegraphics[width=\linewidth]{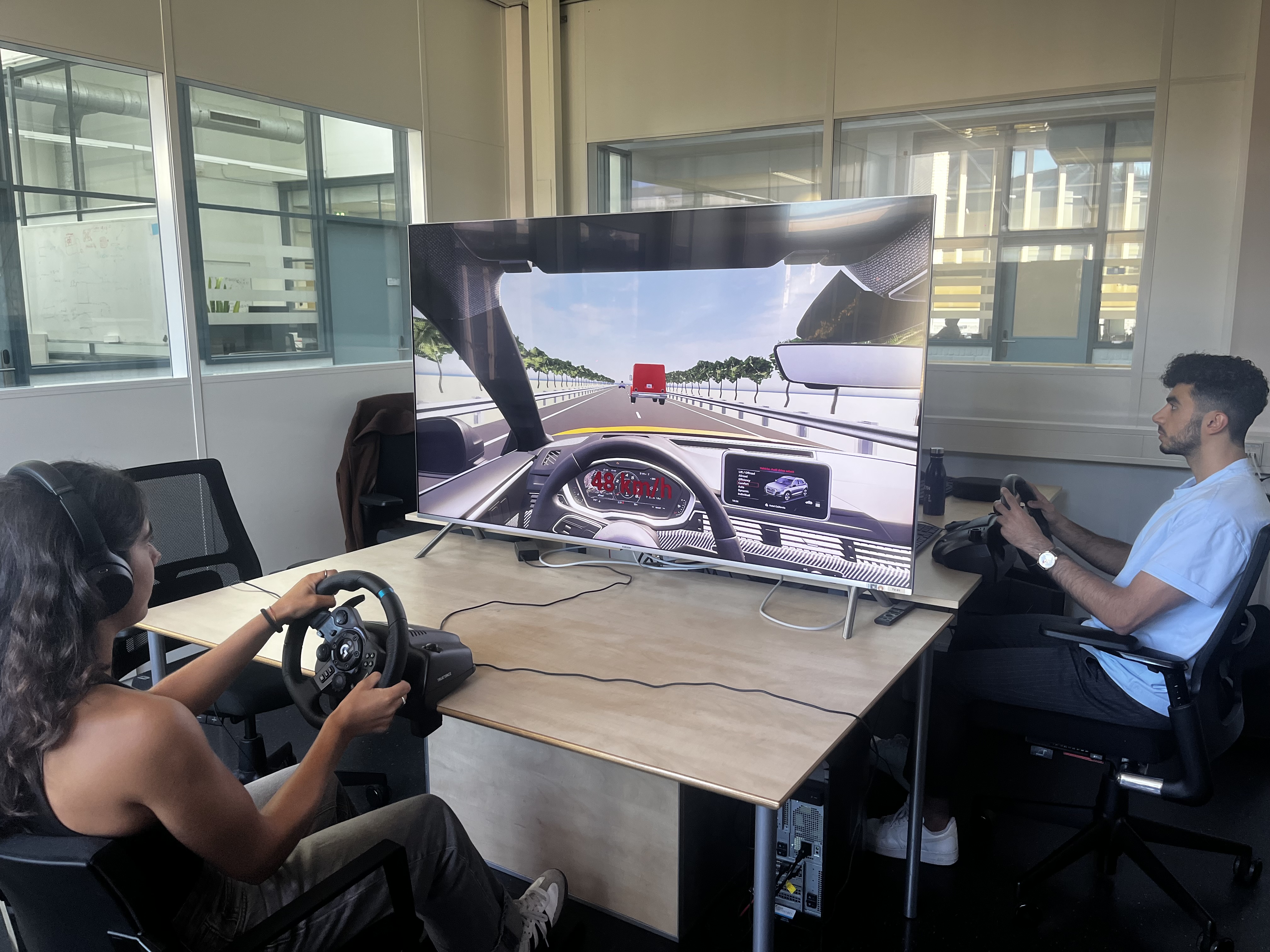}
    \caption{``Reverse Wizard-of-Oz'' experimental setup of the driving simulator. During the sessions involving oncoming ``human-driven'' vehicles, the experimenter (on the right-hand side) pretended to operate an unconnected driving simulator.}
    \label{fig:setup_hardware}
\end{figure}

\begin{figure}[!htbp]
    \centering
    \includegraphics[width=\linewidth]{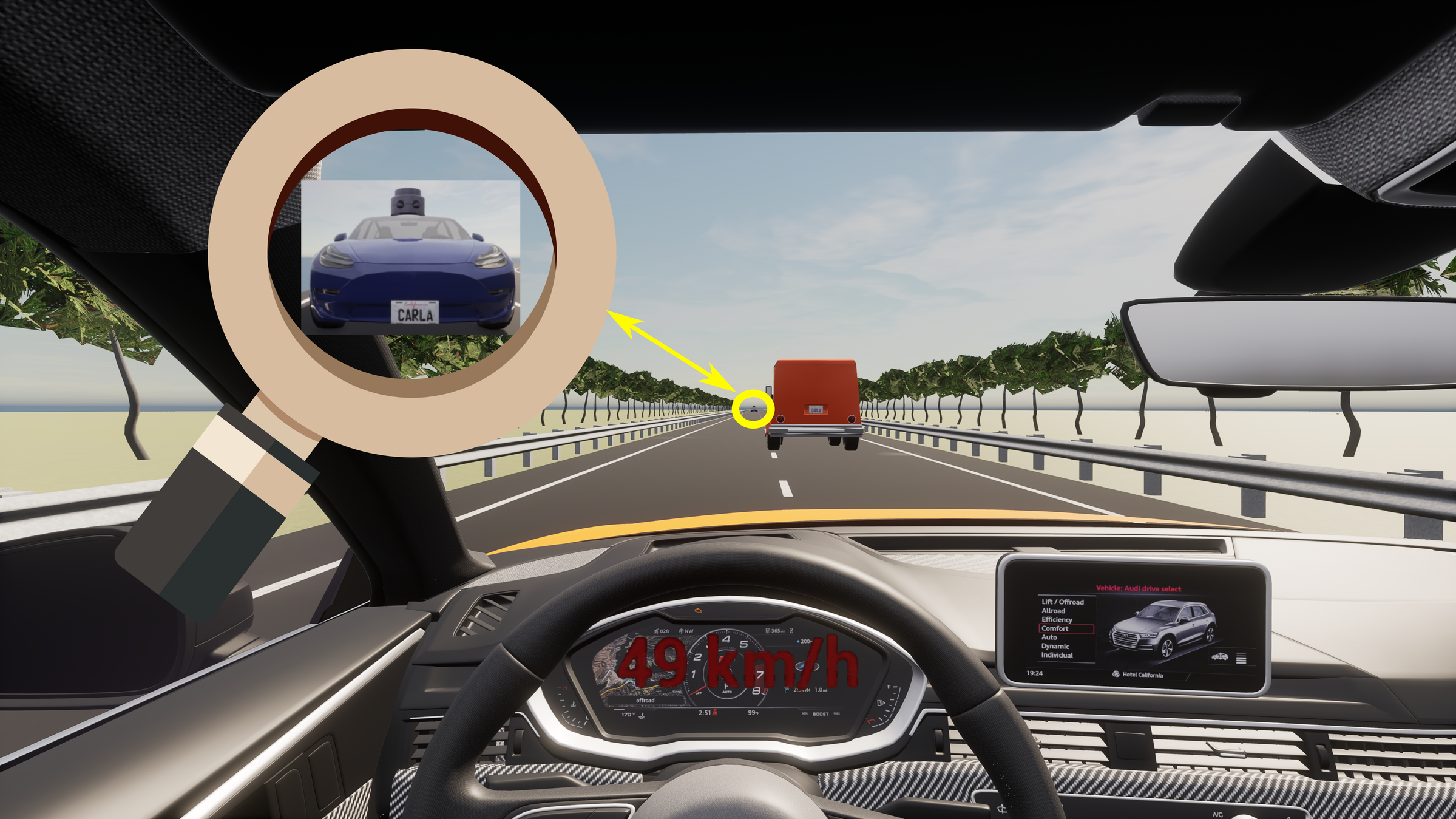}
    \caption{Participants' perspective while performing the task in the driving simulator. As the participant moves to the opposing lane to assess the road situation a decision has to be made to either overtake the lead vehicle (accepting the gap) or stay behind the lead vehicle until the oncoming car passes (rejecting the gap). The inset depicting an oncoming vehicle is for illustrative purposes only and was not present in participants' view during the experiment.}
    \label{fig:setup_view}
\end{figure}

\subsection{Setup}
The experiment utilized a fixed-base driving simulator  (Figure \ref{fig:setup_hardware}). The simulator featured a 65-inch screen and was equipped with a Logitech G923 steering wheel.
The experiment was configured using open-source software frameworks JOAN \cite{beckers2023joan} and CARLA~\cite{dosovitskiy2017carla}. The experimental environment, comprising a straight two-lane rural road, was designed in MathWorks RoadRunner.

\subsection{Experimental design}
\begin{figure}[!thbp]
    \centering
    \includegraphics[width=\linewidth]{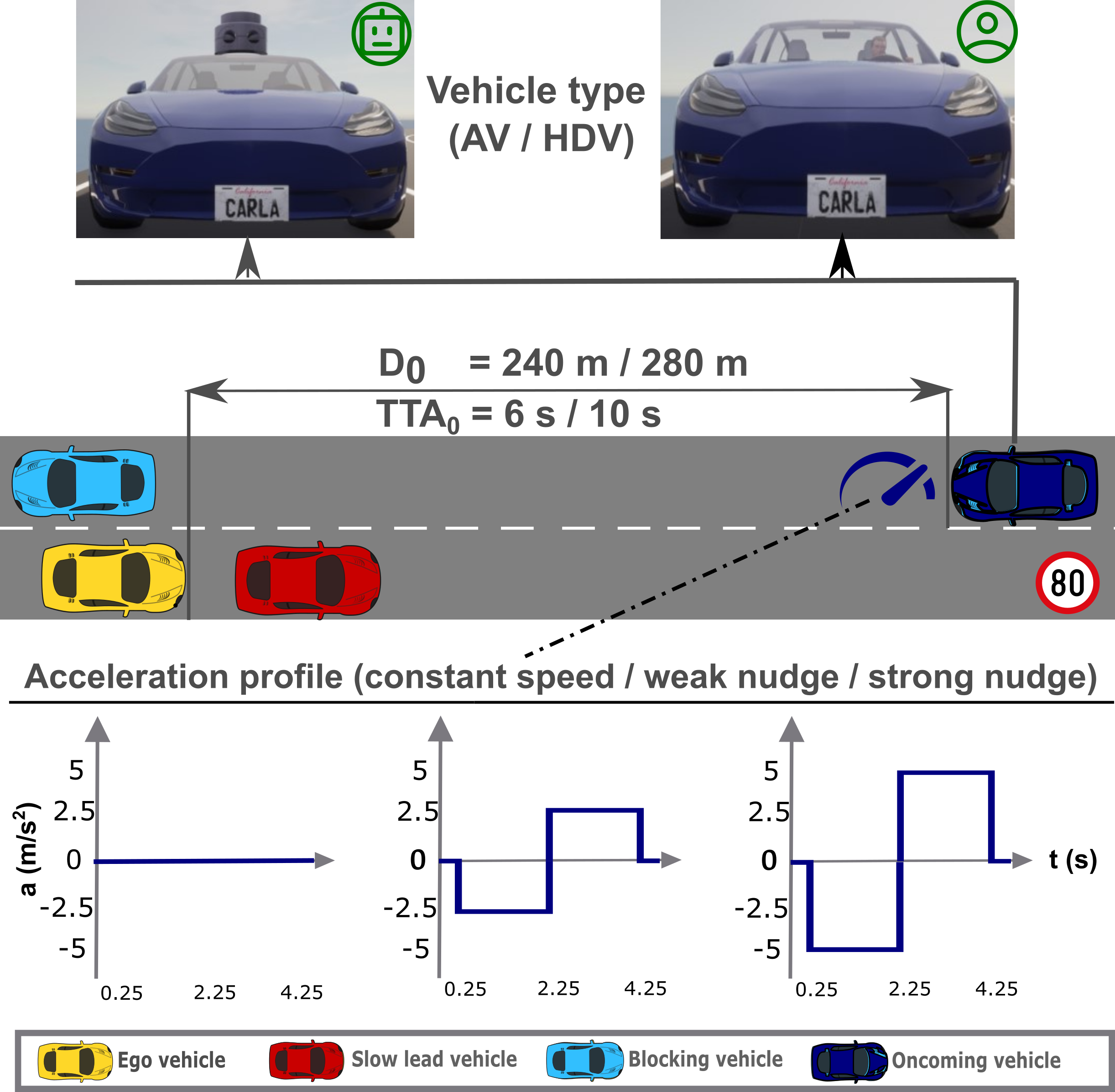}
    \caption{Independent variables manipulated in the experiment: vehicle type, initial distance ($\text{D}_0$), initial time-to-arrival ($\text{TTA}_0$), and acceleration profile of the oncoming vehicle.}
    \label{fig:setup_scheme}
\end{figure}

Participants were instructed to drive in the simulator as they normally would in real life. Participants were informed that oncoming traffic would be present in the opposite lane. Each trial started with three oncoming vehicles with small distances between each other (i.e., a platoon) passing on the opposite lane to block premature overtaking maneuvers. The ego vehicle was set to cruise control until the platoon of vehicles in the opposite lane passed it. After the last vehicle in the platoon passed the ego vehicle, an auditory signal (beep) was delivered through headphones, indicating the start of an overtaking situation (Figure \ref{fig:setup_view}). At this point, participants gained full control over the ego vehicle, which had a headway of approximately 1.5 seconds behind the lead vehicle. Subsequently, to induce a desire to overtake, the lead vehicle speed was gradually reduced from 60 km/h to 45 km/h over 4 seconds. Following a methodology akin to Sevenster et al. \cite{SEVENSTER2023329}, at the time the participant started veering out of their lane, the oncoming vehicle appeared at a given distance ($\text{D}_0$) and time-to-arrival ($\text{TTA}_0$), executing one of the three longitudinal maneuvers (Figure \ref{fig:setup_scheme})). As a result, the participant then had to assess the gap to the oncoming vehicle and make a decision whether to overtake the lead vehicle. 

In our experiment, we kept the lead vehicle's velocity during the overtaking maneuver constant at 45 km/h, while the initial gap to the oncoming vehicle was varied through adjustments in the initial distance (240 and 280 meters) and time-to-arrival (TTA; 6 seconds and 10 seconds) (Figure \ref{fig:setup_scheme}). Given an average speed of 45 km/h for the ego vehicle, the initial velocity of the oncoming vehicle ranged between 40 km/h (low distance, high TTA) and 120 km/h (high distance, low TTA). This setup was chosen to mimic the conditions reasonably encountered on rural roads ~\cite{llorca2016passing, Polus2000EvaluationOT}.  
The behavior of the oncoming vehicle during these interactions was also manipulated: it maintained a constant speed or executed a deceleration of 2.5 $m/s^2$ (weak nudge) or 5 $m/s^2$ (strong nudge) for 2 seconds, after which it accelerated back to its original speed over another 2 seconds. These acceleration profiles were designed to investigate the potential of using brief decelerations as a means of implicit communication by the oncoming AV. Similar nudging maneuvers have previously shown their effectiveness in interactions at intersection crossing~\cite{nudge}, but have not been investigated in the context of overtaking. We posited that varying dynamics over 4 seconds might influence the overtaking decision-making process, considering that such decisions typically span 1 to 3 seconds~\cite{SEVENSTER2023329}. The exact deceleration values of 2.5 $m/s^2$ and 5 $m/s^2$ were selected based on pilot experiments.

Additionally, we alternated the oncoming vehicle type (AV and HDV) over two sessions throughout the experiment. In the session featuring an oncoming HDV, we employed a ``reverse Wizard-of-Oz'' setup where the experimenter was pretending to operate another driving simulator (Figure \ref{fig:setup_hardware}). This setup created the illusion that the oncoming HDV was human-controlled. Furthermore, the HDV had an animated driver and no LiDAR (Figure \ref{fig:setup_scheme}). 

The experiment thus followed a within-participant design with a 2 x 2 x 2 x 3 factorial structure, with four independent variables: the initial distance between the ego vehicle and the oncoming vehicle (240m or 280m), the initial TTA (6 s or 10 s), the oncoming vehicle type (AV or HDV), and the acceleration profile of the oncoming vehicle (constant speed, weak nudge, or strong nudge) (Figure \ref{fig:setup_scheme}). 

In total, there were 24 unique conditions. Two additional conditions were included in which the gap was very small ($D_0\in\{70,150\}m$, $\text{TTA}_0$ = 2 s, constant speed). These conditions were added to encourage participants to perform a careful assessment of the road situation and not simply overtake in every trial. The data from these conditions was excluded from the analysis.

To familiarize themselves with the driving equipment and task, participants were asked to perform 5 to 10 practice trials before the start of the experiment, ensuring their comfort with the experimental procedure. Then, 26 conditions were repeated five times, randomly shuffled and split evenly into two sessions based on vehicle type, resulting in a total of 130 trials. The session order was randomized between participants. To maintain participant concentration, a brief off-screen task followed after every 13 trials. Each session lasted approximately 45 minutes, with a 15-minute break between the sessions. After the second session, participants answered questions regarding perceived safety and the behavior of the oncoming vehicle in a post-experiment questionnaire. Overall, the experiment recorded a total of about 3600 overtaking gap acceptance decisions.

\subsection{Recorded data and metrics}
The data, including trajectories and velocities of the ego vehicle and the oncoming vehicle, was captured at a rate of 100 Hz. 

From this data, we extracted the two main dependent variables: the decision outcome (Overtake (gap accepted) or Stay (gap rejected)) and response time (the duration of the gap acceptance decision-making process).

The decision was determined by checking whether the ego vehicle returned to its lane behind the lead vehicle (indicating gap rejection) or not (indicating gap acceptance). The response time was determined as the difference between $t_{\text{end}}$ (the timestamp corresponding to the end of the decision-making process) and $t_{\text{start}}$ (the time the decision-making process started). 

\begin{figure*}
    \centering
    \includegraphics[width=0.8\linewidth]{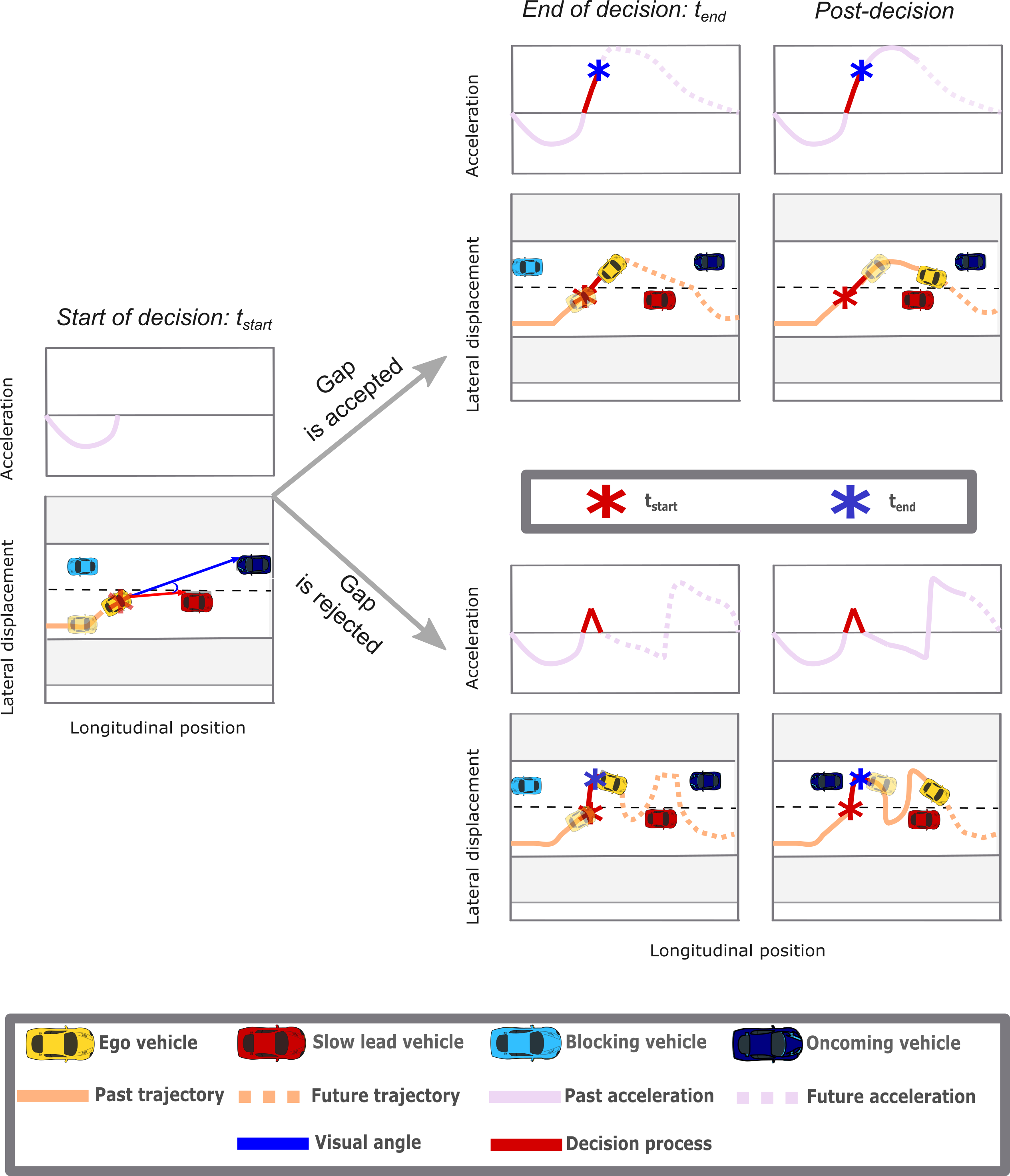}
    \caption{Response time measurement in rejected gaps \cite{SEVENSTER2023329} and our proposed measurement method in accepted gaps. The moment the oncoming vehicle appeared in participants' field of view was denoted as the start of the decision-making process ($t_{\text{start}}$). The end of the decision-making process ($t_{\text{end}}$) was defined as the moment the ego vehicle acceleration reached a pre-defined threshold of 3 $m/s^2$ (in accepted gaps) or the moment when the ego vehicle started returning to its original lane after veering into the opposite lane to assess the gap (in rejected gaps). }
    \label{fig:RT}
\end{figure*}

The start of the decision-making process was determined as the moment the oncoming vehicle appeared in the field of view of the ego vehicle. The moment the decision-making process ended was quantified differently depending on whether the gap was rejected or accepted (Figure \ref{fig:RT}). 

For rejected gaps, $t_{\text{end}}$ was calculated following the method proposed by Sevenster et al.~\cite{SEVENSTER2023329} (Figure \ref{fig:RT}): the decision was considered finalized at the moment the ego vehicle started returning to its original lane after veering into the opposite lane to assess the gap. 

For accepted gaps, we denoted the end of the decision-making process as the point where the acceleration of the ego vehicle surpassed a predetermined threshold of 3 $m/s^2$ (Figure \ref{fig:RT}). We justified this choice by the typical behavior observed in this experiment, where participants generally refrained from accelerating while assessing oncoming traffic due to their proximity to the slow-moving vehicle ahead.

\subsection{Exclusion criteria} \label{exclusion}
We excluded trials in which the overtaking decision could not be determined due to vehicle collisions ($n=19$) and instances where the response time in accepted decisions could not be accurately measured ($n=75$) due to missing acceleration data. Additionally, we identified and removed instances of unrealistic response times, both excessively short ($< 0.5 s$, $n=225$) and exceptionally long ($> 4 s$, $n=29$). These exclusions were made in the context of statistical analyses involving response times and cognitive modeling but not in the statistical analyses of decision outcomes.

In total, our analyses were based on 3438 overtaking maneuvers to assess decision outcomes and 3184 decisions for analyzing response times and cognitive modeling.

\subsection{Data analysis}
We conducted statistical analyses using mixed-effect regressions for decision outcomes (logistic) and response times (linear) in \textit{pymer4}~\cite{jolly2018pymer4}. Dummy coding was employed for vehicle type and acceleration profile, using AV and constant speed as the respective reference groups. To address variations in baseline values of dependent variables across individuals, we included the vehicle type per participant as a random slope as well as random intercept per participant in all regression models.

For statistical analyses, we standardized all continuous variables (initial distance $D_{0}$, initial time-to-arrival $\text{TTA}_0$, and initial ego vehicle velocity $v_0^{ego}$) through z-scoring. This standardization allowed us to interpret the coefficients ($\beta$) for each independent variable in terms of their relative contributions to the dependent variable. Additionally, we dichotomized the values of initial ego vehicle velocities into two equally-sized clusters: low ($v_0<13.8$m/s) and high ($v_0>13.8$m/s) velocities. This was done for visualization and to enable inclusion of the initial velocity as a factor in the drift-diffusion models; for statistical analyses, the original values of $v_0$ were used. 

In the case of the response time regression, we computed the Type-III sum-of-squares ANOVA table, utilizing the Satterthwaite approximation for degrees of freedom. To account for multiple comparisons in both decision and response time regression analyses, particularly concerning the acceleration profiles, we adjusted p-values using the Tukey method.

\subsection{Implementation and fitting of cognitive models}
Cognitive models were implemented using \textit{pyddm} \cite{shinn} and fitted to the data using the differential evolution optimization technique with Bayesian information criterion as a loss function. 

We aimed to assess whether our models effectively captured the general trends in the behavior of our participants, rather than explaining individual differences. Hence, we fitted the models to the ``average'' participant by aggregating all the data. Furthermore, since there was no evidence for differences in participants' overtaking behavior between oncoming AVs and HDVs, as well as no order effects, we excluded vehicle type and session order factors from all cognitive models.

\section{Results} \label{results}

\begin{table}[]
    \caption{Coefficients of the mixed-effect logistic regression describing the final decision as a function of z-scored variables $D_{0}$, $TTA_{0}$ and $v_0^{ego}$, acceleration profile, vehicle type, and session order. The vehicle type per participant ID was included as a random slope.}
    \centering
    \begin{tabular}{r|r|r|r|r} \hline
         & \textit{$\beta$} & SE & \textit{z} & \textit{p} \\ \hline
        (Intercept) & -0.25 & 0.20 & -1.2 & 0.23 \\
        $D_{0}$ & 0.96 & 0.04 & 21.7 & $< 0.001$ \\
        $TTA_{0}$ & 0.58 & 0.04 & 13.6 & $< 0.001$ \\
        $v_0^{ego}$ & 0.50 & 0.06 & 9.03 & $< 0.001$ \\
        Acceleration profile ``weak nudge" & 0.06 & 0.10 & 0.55 & 0.58 \\
        Acceleration profile ``weak nudge" & 0.07 & 0.10 & 0.65 & 0.51 \\
        Vehicle type HDV & 0.12 & 0.12 & 1.06 & 0.29 \\
        Session order second & -0.12 & 0.12 & -1.03 & 0.30 \\ \hline
    \end{tabular}

    \label{tab:coeff}
\end{table}

\begin{figure*}[!htbp] % Use figure* for a full-width (two-column) figure

  \begin{subfigure}{\textwidth} % Adjust the width as needed
    \centering
    \includegraphics[width=\linewidth]{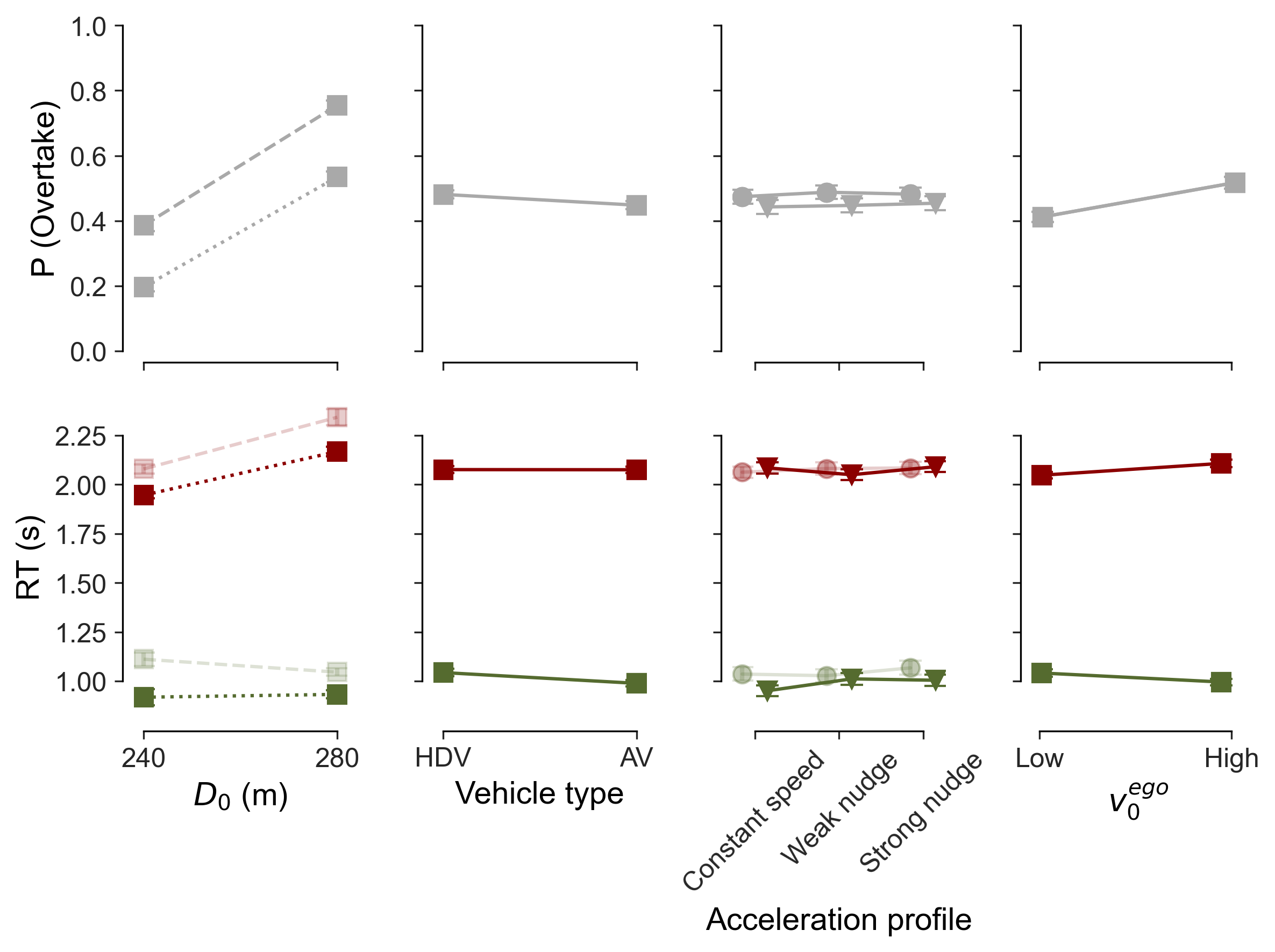}
  \end{subfigure}%

  \begin{subfigure}{\textwidth} % Adjust the width as needed
    \centering
    \includegraphics[width=\linewidth]{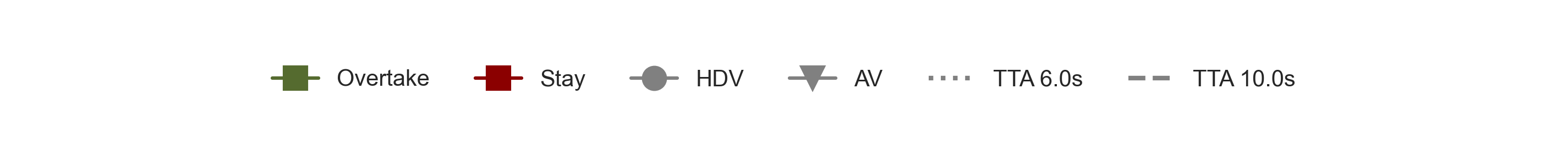}
  \end{subfigure}%
  
  \caption{Overview of the average participant's behavior in overtaking decisions. For visualization purposes, the initial velocity of the ego vehicle $v_0^{ego}$ was dichotomized into low and high groups (with the cutoff value of $13.8m/s$). Error bars denoting 95\% confidence intervals are narrower than the marker size in some panels and therefore not visible.}
  \label{exp_results}
\end{figure*}

\subsection{Decision outcomes}
The probability of accepting the gap (i.e., the Overtake decision) was positively affected by the initial distance $D_0$ and initial time gap $\text{TTA}_0$ to the oncoming vehicle, as well as the initial velocity of the ego vehicle $v_0^{ego}$ (Table \ref{tab:coeff}, Figure~\ref{exp_results}). 

In line with our hypothesis, post-hoc comparisons showed no evidence that the probability of overtaking differed between vehicle types AV vs. HDV ($\Delta$ = -0.12, \textit{z} = -1.06, \textit{p} = 0.29); this was the case in both the first ($p = 0.52$) and the second session ($p = 0.21$). Despite no evidence of the overall effect on the group level however, we did observe substantial individual differences (see online supplementary information).

Contrary to our hypothesis, there was no evidence of differences in overtaking probability across different acceleration profiles: ``constant speed" vs ``weak nudge" ($\Delta$ = -0.071, \textit{z} = -0.69, $p = 0.77$), ``constant speed" vs ``weak nudge" ($\Delta$ = -0.057, \textit{z} = -0.56, $p = 0.84$), and ``weak nudge" vs ``weak nudge" conditions ($\Delta$ = -0.014, \textit{z} = -0.14, $p = 0.99$). 

\subsection{Response times}
We observed significant influences of the decision outcome, initial distance $D_{0}$, and initial time gap $TTA_{0}$ on response times (Table \ref{tab:anova}, Figure \ref{exp_results}). Post-hoc comparisons showed that Overtake responses were faster than Stay responses ($\Delta$ = -1.15 s, \textit{t} = -59.9, $\textit{p} < 0.001$). Initial distance $D_{0}$ positively affected Stay response times (\textit{b} = 0.13, \textit{t} = 11.4, $\textit{p} < 0.001$), and, marginally, Overtake response times ($\textit{p} = 0.045$). Both Overtake and Stay responses times increased with $TTA_{0}$; Overtake--- \textit{b} = 0.058, \textit{t} = 4.46, $\textit{p} < 0.001$, and Stay--- \textit{b} = 0.085, \textit{t} = 7.72, $\textit{p} < 0.001$. Response times in rejected gaps increased with initial ego velocity ($b = 0.04, \textit{p} < 0.001$), although the effect was small (Figure~\ref{exp_results}). At the same time, initial ego vehicle velocity $v_0^{ego}$ did not substantially affect Overtake response times ($p = 0.053$). 

Our analysis of response times did not reveal evidence for main effects of vehicle type ($p = 0.5$) or acceleration profile ($p=0.35$), although we did observe a marginally significant interaction between vehicle type and decision ($p=0.046$). Post-hoc tests revealed no evidence for differences in Overtake response times between conditions ``constant speed" vs ``weak nudge"($\Delta$ = -0.01 s, \textit{t} = -0.29, $p = 0.95$), ``constant speed" vs ``weak nudge" ($\Delta$ = -0.044 s, \textit{t} = -1.27, $p = 0.41$), and ``weak nudge" vs ``weak nudge" ($\Delta$ = 0.034 s, \textit{t} = 1.01, $p = 0.57$). Likewise, no evidence for differences was found in Stay response times:  ``constant speed" vs ``weak nudge" ($\Delta$ = -0.002 s, \textit{t} = -0.064, $p \approx 1.0$), ``constant speed" vs ``weak nudge" ($\Delta$ = -0.020 s, \textit{t} = -0.76, $p = 0.73$), and ``weak nudge" vs ``weak nudge" $\Delta$ = 0.019 s, \textit{t} = 0.69, $p = 0.77$) conditions.

\begin{table}[!htbp]
    \caption{ANOVA table based on the mixed-effect linear regression describing response time as a function of decision, acceleration profile, vehicle type, session order, and z-scored variables $D_{0}$, $TTA_{0}$, and $v_0^{ego}$.}
    \centering
    \begin{tabular}{r|r|r|r|r|r} \hline
         & SS & MS & df & \textit{F} & \textit{p} \\ \hline
        Decision & 640 & 640 & 1 & 3525 & $< 0.001$ \\
        $D_{0}$ & 6.56 & 6.56 & 1 & 36.1 & $< 0.001$ \\
        $TTA_{0}$ & 14.0 & 14.0 & 1 & 77.2 & $< 0.001$ \\
        $v_0^{ego}$ &  0.20 & 0.20 & 1 & 1.11 & 0.29 \\
        Acceleration profile & 0.38 & 0.19 & 2 & 1.04 & 0.35 \\
        Vehicle type & 0.09 &  0.09 & 1 & 0.48 & 0.50 \\
        Session order & 0.12 &  0.12 & 1 & 0.66 & 0.42 \\
        Decision:$D_{0}$ & 13.2 & 13.2 & 1 & 72.7 & $< 0.001$ \\
        Decision:$TTA_{0}$ & 0.50 & 0.56 & 1 & 2.75 & 0.1 \\
        Decision:$v_0^{ego}$ &  2.62 & 2.62 & 1 & 14.4 & $< 0.001$ \\
        Decision:Acceleration profile & 0.18 & 0.09 & 2 & 0.51 & 0.60 \\
        Decision:Vehicle type & 0.72 &  0.72 & 1 & 3.99 & 0.046 \\
        Decision:Session order & 0.07 &  0.07 & 1 & 0.37 & 0.54 \\
        \hline
    \end{tabular}

    \label{tab:anova}
\end{table}

\subsection{Post-experiment questionnaire}
In the post-experiment questionnaire, participants reported a similar sense of safe interactions in the two sessions (AV: mean = 3.9, SD = 0.76 vs HDV: mean = 3.9, SD = 0.64,  \textit{t} = -0.25, $p = 0.83$). Lastly,  despite the lack of differences in the actual behavior of AVs and HDVs, participants were moderately certain that the AV and the HDV behaved differently (mean = 3.1, SD = 1.3).

\subsection{Summary} \label{resultscon}
Based on the experimental findings, we can conclude the following.
\begin{itemize}
    \item Initial distance, initial TTA, and initial ego vehicle velocity positively affected overtaking probability 
    \item Initial TTA positively affected the response times for both decisions, while initial distance positively affected response times for Stay but not Overtake decisions. 
    \item No evidence of a difference in overtaking probability between the oncoming vehicle types AV and HDV was found.
    \item There was no evidence that weak and strong nudges impact overtaking probability compared to the constant-speed oncoming vehicle.
    \item There was no evidence that the oncoming vehicle type or its acceleration profile affected the response times.
\end{itemize}

\section{Cognitive process modeling} \label{cognitive}
\subsection{Basic drift-diffusion model and its applications to traffic}
We utilized the drift-diffusion modeling (DDM) framework~\cite{Ratcliff1978ATO} to describe participants' decision-making processes in our experiment. According to the DDM, decision making is an ongoing process of accumulating relevant perceptual information over time. This process is noisy (reflecting the assumption that the perceived information is perceived and processed imperfectly) and bounded, such that a decision is made when a certain amount of evidence is accumulated.

Mathematically, the rate of evidence accumulation is denoted as the drift rate $s(t)$, while the random factor (diffusion) is characterized as a stochastic variable $\epsilon(t)$ (white noise). The momentary evidence $x$ favoring one alternative emerges from integrating both drift and diffusion:
\begin{equation}
\frac{dx}{dt}=s(t) + \epsilon(t), \quad x(t_0) = Z.
\label{eq:ddm}
\end{equation}
% (Eq. \eqref{eq:ddm}). 
This continuous process, beginning at position Z---which indicates its proximity to a boundary---is bounded and concludes when the evidence favoring one alternative reaches a predetermined boundary  ($x = \pm b(t)$). Finally, DDM also incorporates non-decision time, which accounts for the duration of cognitive processes not directly related to decision-making, such as perceptual and motor delays.

Despite its computational simplicity, DDMs have proven highly effective in modeling and comprehending a wide array of decision-making processes, encompassing choice behavior and response times in experimental investigations~\cite{RATCLIFF2016260,evans2019evidence}. More recently, DDMs have also been successfully applied to traffic-related decision processes in the presence of dynamically changing evidence such as gap acceptance in pedestrian crossings~\cite{Pekkanen2022} and left turns at unprotected intersections~\cite{ZgonnikovAbbinkMarkkula, nudge, bontje_are_2024}. These studies emphasized that the drift rate $s(t)$ and possibly boundaries $b(t)$ need to be contingent on dynamically evolving gap sizes. 

However, in contrast to pedestrian crossing and left-turn decisions, in overtaking maneuvers, the decision maker is moving with a high velocity while making the decision. The influence of this initial velocity on decision outcomes was evident in the overtaking experiment conducted by Sevenster et al.~\cite{SEVENSTER2023329}. In particular, they found that initial velocity positively influenced gap acceptance probability while negatively affecting response times in accepted gaps. Drawing upon the data from Sevenster et al.~\cite{SEVENSTER2023329}, Mohammad et al.~\cite{mohammad2023cognitive} explored various versions of the DDM where the initial velocity was integrated into different components of the model: drift rate, decision boundary, and the starting point $Z$. The simplest model capable of effectively capturing all qualitative patterns in their used dataset included a drift rate dependent on both distance and TTA, the boundaries collapsing as distance and TTA decreased, and, importantly, the starting point $Z$ dependent on the initial velocity of the ego vehicle.

\begin{figure}[ht]
    \centering
    \includegraphics[width=\linewidth]{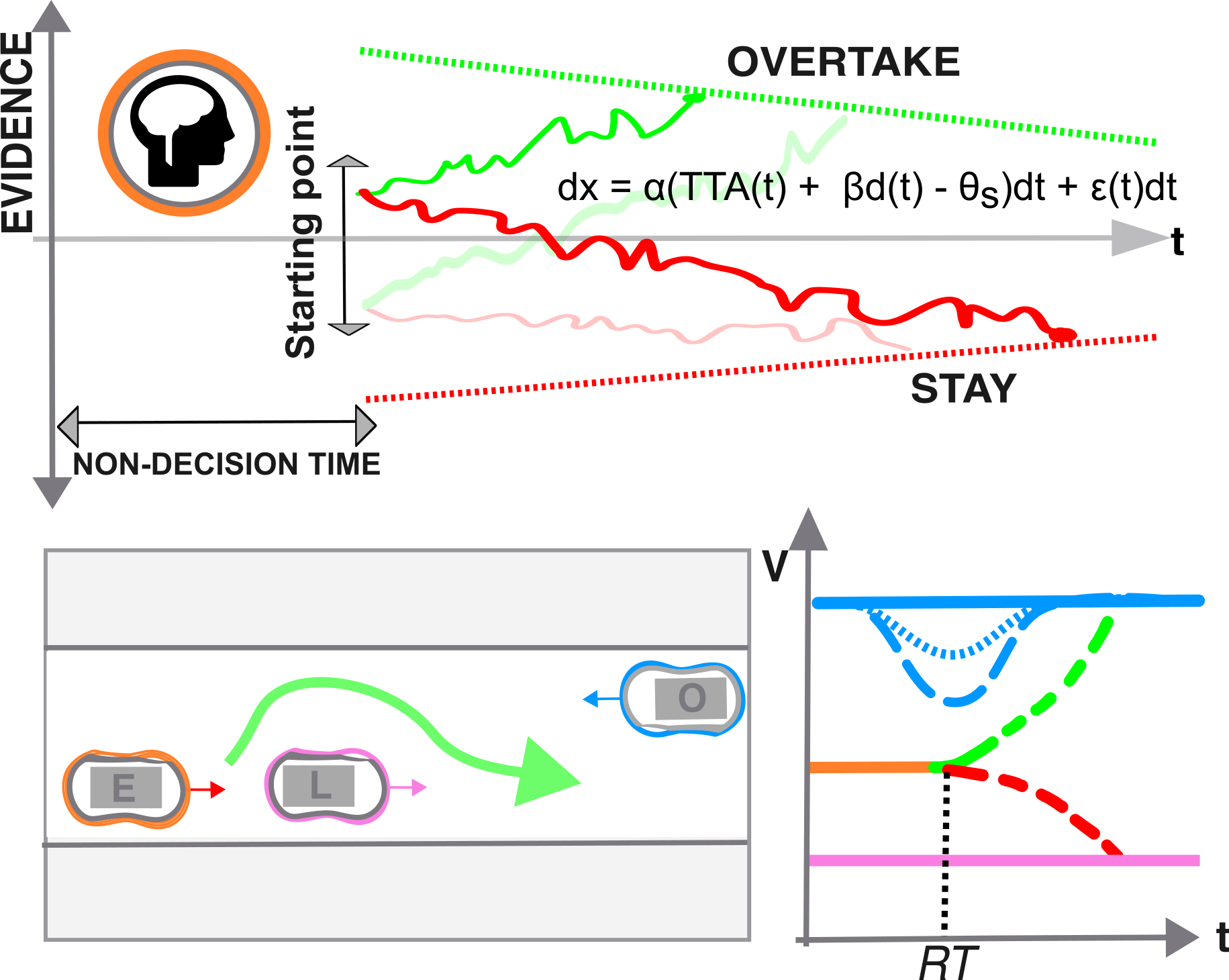}
    \caption{Drift-diffusion model of gap acceptance in an overtaking scenario. Pink represents the lead vehicle, blue is the oncoming vehicle, and orange is the human-driven ego vehicle. Red indicates staying in the lane, and green represents overtaking, while blue velocity curves depict different dynamics of the oncoming vehicle.}
    \label{fig:cognitive-overtake}
\end{figure}

\subsection{Candidate drift-diffusion models for dynamic overtaking scenarios}
Although previous studies provided early evidence that DDMs can be applied to overtaking decisions, these studies were limited to situations with a constant-acceleration oncoming vehicle~\cite{SEVENSTER2023329, mohammad2023cognitive}. Furthermore, our experiment differed substantially from the one conducted by Sevenster et al. \cite{SEVENSTER2023329} in other aspects, with variations in the initial distance (160m and 220m vs. 240m and 280m) and additional controlled variable (initial TTA). Thus, to shed light on cognitive processes underlying the overtaking decisions and response times of our participants, we explored multiple models to find the one that fits our dataset best. To do this, we re-evaluated the four main components of the DDM framework used by Mohammad et al. \cite{mohammad2023cognitive} and proposed four potential models to explain our experimental results (Table~\ref{tab:modelvariations}). The model originally highlighted by Mohammad et al. as the one most consistent with the data of Sevenster et al. will be referred to here as the baseline model (Figure~\ref{fig:cognitive-overtake}).

\subsubsection*{Non-decision time}
For all of our models, the non-decision time is assumed to vary randomly across trials, following a normal distribution:

\begin{equation}
 t^{ND} \in \mathcal{N}(\mu_{ND},\sigma_{ND}),\quad \mu_{ND}>0,\, \sigma_{ND}>0.   
 \label{eq:ndt}
\end{equation}

\subsubsection*{Drift rate}
The drift rate $s(t)$ in all our models is determined by parameters $\alpha>0$, $\beta>0$, and $\theta_s>0$, and is a measure of relative evidence $x$ favoring either the Overtake or Stay decision at any given moment $t$:
\begin{equation}
 s(t) = \alpha(TTA(t) + \beta d(t) - \theta_s) \label{eq:drift}   
\end{equation}

The size of the time and distance gap (combined with the weighting factor $\beta$) between the ego vehicle and the oncoming vehicle, relative to the critical gap value $\theta_s$ determines the drift rate's direction. Specifically, if the combined gap is larger than $\theta_s$, the drift rate is positive, suggesting a higher likelihood of the decision-maker leaning towards the Overtake decision. Conversely, a gap smaller than $\theta_s$ leads to a negative drift rate, indicating a greater probability of choosing the Stay decision. The gap itself is dynamic and can increase during the decision-making process, for example, when the oncoming vehicle decelerates. This change in gap size directly impacts the drift rate by affecting how the current gap compares to the critical threshold $\theta_s$. 

\subsubsection*{Decision boundary}
The accumulation process ends upon reaching either boundary (positive or negative) with the height of each boundary representing how much evidence is required for choosing the respective alternative.

Intuitively, with lower values of $TTA(t)$ and $d(t)$, the decision-maker might experience a stronger sense of urgency to make a decision, which can potentially be reflected in the boundary $b(t)$ decreasing with gap size. Such collapsing boundaries have been shown to be beneficial for describing left-turn gap acceptance~\cite{ZgonnikovAbbinkMarkkula,bontje_are_2024}, although constant-boundary models can also outperform models with collapsing boundaries~\cite{nudge}. Since initial TTA did affect response times in our data (Figure \ref{exp_results}), we deemed it worth examining whether decreasing TTA urges the driver to make a decision faster. Hence, we tested two candidate assumptions: boundaries constant over time 
\begin{equation}
b(t) = \pm B,
\label{eq:bound_constant}
\end{equation}
and boundaries exponentially collapsing based on kinematic variables $d(t)$ and $TTA(t)$ in relation to a critical value $\theta_s$, with a steepness parameter k and starting from an initial boundary height $b_{0}$
\begin{equation}
b(t) = \pm \frac{b_0}{1+e^{-k(TTA(t) + \beta d(t) -\theta_s)}}.
\label{eq:bound_drift}
\end{equation}

\begin{table*}[t]
    \centering
    \caption{Four tested variations of the generalized drift-diffusion model~\eqref{eq:ddm} with varying boundary functions (Eq.~\eqref{eq:bound_constant}~\eqref{eq:bound_drift}) and starting point functions (Eq.~\eqref{eq:bias_constant}~\eqref{eq:bias_vel}). All four models used the same drift rate (Eq.~\eqref{eq:drift}) and non-decision time (Eq.~\eqref{eq:ndt}). The number of parameters in the last column thus includes not only the decision boundary and starting point parameters, but also the drift rate parameters $\alpha$, $\beta$, $\theta_s$ and the non-decision time parameters $\mu_{ND}$, $\sigma_{ND}$.}
\begin{tabular}{lclclc}
Model &  Decision boundary $b(t)$ & Eq. & Initial bias $ -b(t_0) < Z < b(t_0)$ & Eq. & \# parameters  \\
M1 & $ \pm \frac{b_0}{1+e^{-k(TTA(t) + \beta d(t) -\theta_s)}}$ &\eqref{eq:bound_drift} & $C_z$ & \eqref{eq:bias_constant} & 8\\
M2 & $ \pm B$ &\eqref{eq:bound_constant} & $C_z$ & \eqref{eq:bias_constant} & 7\\
M3 (baseline \cite{mohammad2023cognitive}) &$ \pm \frac{b_0}{1+e^{-k(TTA(t) + \beta d(t) -\theta_s)}}$ &\eqref{eq:bound_drift} & $\frac{2b(t_0)}{1+e^{-b_z(v^0_{ego}-\theta_z)}} - b(t_0)$ & \eqref{eq:bias_vel} & 9\\
M4 & $ \pm B$ &\eqref{eq:bound_constant} & $\frac{2b(t_0)}{1+e^{-b_z(v^0_{ego}-\theta_z)}} - b(t_0)$ & \eqref{eq:bias_vel} & 8\\
\end{tabular}
\label{tab:modelvariations}
\end{table*}

\begin{figure*} % Use figure* for a full-width (two-column) figure
% \begin{subfigure}{\textwidth}
    \centering
    \includegraphics[width=\linewidth]{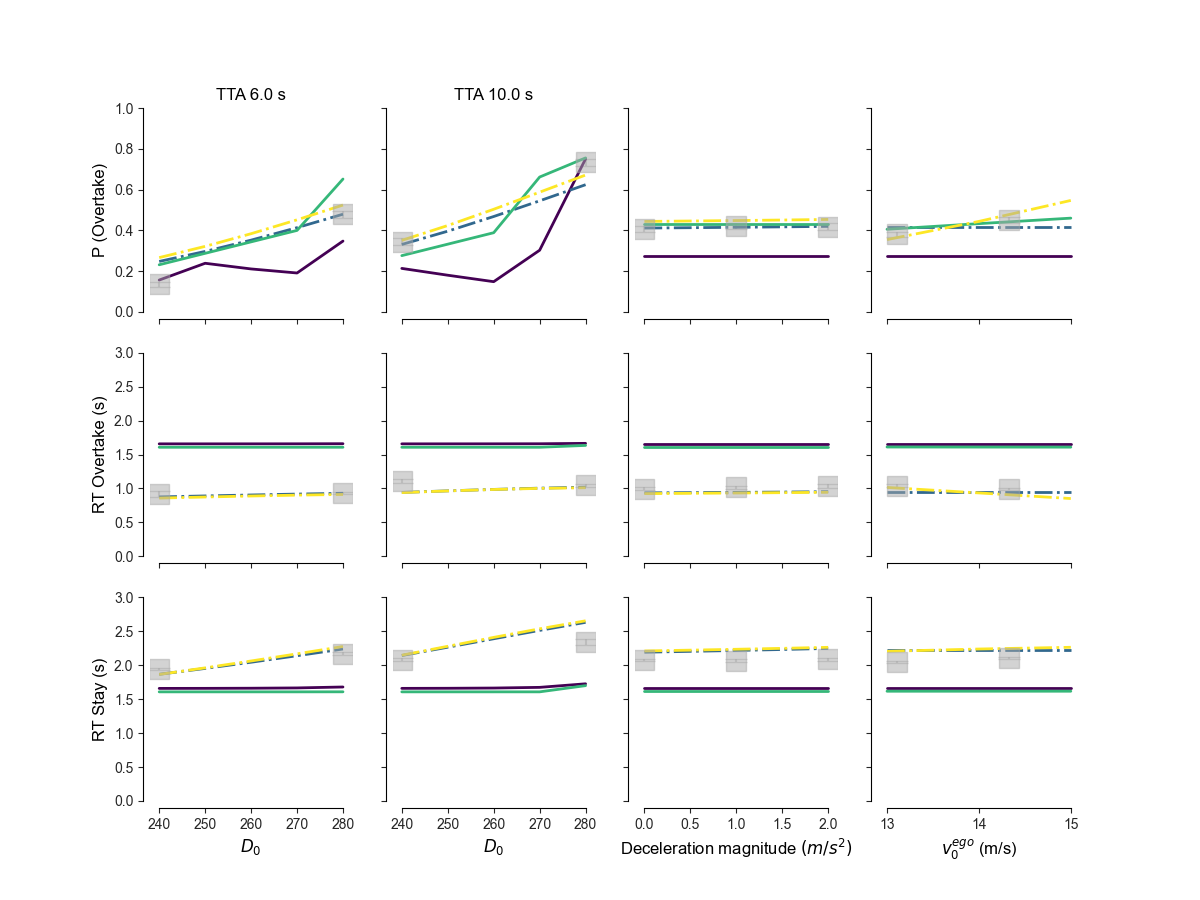}
% \end{subfigure}%
% \begin{subfigure}{\textwidth}
%     \centering
    \includegraphics[width=\linewidth]{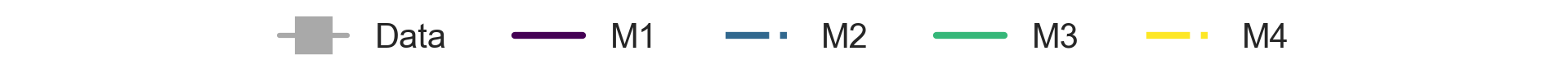}
% \end{subfigure}%
\caption{Simulated model results compared to the experimental data to show the effect of distance and TTA, deceleration magnitudes of acceleration profiles, and the initial ego vehicle velocity on gap acceptance behavior. The error bars represent the standard error of the mean. }
\label{fig:results}
\end{figure*}

\begin{table*}[!t]
    \centering
    \caption{Qualitative assessment of candidate drift-diffusion models according to the experimental findings.}
    \begin{tabular}{l|cccc}
        \multicolumn{1}{c|}{Finding} & M1 & M2 & M3 & M4 \\
        \hline
        The probability of accepting the gap increases with the initial distance to the oncoming vehicle. & \textbf{X} & \Checkmark  & \Checkmark & \Checkmark \\
        The probability of accepting the gap increases with the initial TTA to the oncoming vehicle. & \textbf{X} & \Checkmark & \Checkmark &\Checkmark  \\
        The probability of accepting the gap increases with the initial velocity of the ego vehicle. & \textbf{X} & \textbf{X}  & \Checkmark  & \Checkmark  \\  
        The probability of accepting the gap is not substantially affected by the acceleration profile of the oncoming vehicle. & \Checkmark & \Checkmark & \Checkmark & \Checkmark \\
        Response times in rejected gaps are higher than in accepted gaps & \textbf{X} & \Checkmark & \textbf{X} & \Checkmark \\
        Response times in accepted gaps are not substantially affected by the initial distance to the oncoming vehicle. & \Checkmark & \Checkmark & \Checkmark & \Checkmark \\
        Response times in rejected gaps increase with the initial distance to the oncoming vehicle. & \textbf{X} & \Checkmark & \textbf{X} & \Checkmark \\
        Response times increase with the initial TTA to the oncoming vehicle. & \textbf{X} & \Checkmark & \textbf{X} & \Checkmark \\
        Response times are not substantially affected by the initial velocity of the ego vehicle. & \Checkmark & \Checkmark &  \Checkmark & \Checkmark \\
        Response times are not substantially affected by the acceleration profile of the oncoming vehicle.& \Checkmark & \Checkmark &  \Checkmark & \Checkmark \\
        \hline
        \textbf{Total} & \textbf{4/10} & \textbf{9/10} & \textbf{7/10} & \textbf{10/10} 
    \end{tabular}
    \label{tab:model_assessments}
\end{table*}

\subsubsection*{Starting point}

Similar to Sevenster et al.~\cite{SEVENSTER2023329}, our experimental results highlighted an effect of the initial velocity of the ego vehicle on decision outcomes and response times (Figure~\ref{fig:results}). Therefore, similar to Mohammad et al.~\cite{mohammad2023cognitive}, we tested two possible variations: fixed starting point 
\begin{equation}
Z = C_z 
\label{eq:bias_constant}    
\end{equation}
and the starting point dependent on the initial velocity of the ego vehicle 

\begin{equation}
 Z = \frac{2b(t_0)}{1+e^{-b_z(v^0_{ego}-\theta_z)}} - b(t_0). \label{eq:bias_vel} 
\end{equation}

Here, a value of $Z < 0$ indicates an initial bias towards the Stay decision, while $Z > 0$ indicates a bias towards the Overtake decision. This bias can be represented by a constant value $C_z$ or can vary based on the initial velocity $v^0_{ego}$ in relation to a critical parameter $\theta_z$. In the latter case, relatively higher and lower initial speeds correspond to a bias toward the Overtake and Stay decisions, respectively. The parameter $b_{z}$ quantifies the strength of this speed-related effect on the starting point. Furthermore, the starting point is constrained within the initial limits of the boundaries $\pm b_{t_{0}}$. 

The four model variants we tested all have the same non-decision time and drift rate components, but differ in their assumptions on decision boundary and starting point (Table \ref{tab:modelvariations}).

\subsection{Model fitting results}
We found that the four tested models differed substantially in their qualitative alignment with the observed behavior of the average participant (Figure~\ref{fig:results}, Table~\ref{tab:model_assessments}). 
All models exhibited consistent behavior regarding the probability of overtaking not varying across various acceleration profiles with magnitudes of deceleration nudges ranging from 0 (constant speed) to 5 $m/s^2$ (strong nudge). However, models with a constant starting point (M1 and M2) failed to account for the effect of initial velocity on gap acceptance probability. Conversely, models featuring an initial velocity-dependent initial bias (M3 and M4) successfully captured this effect.

Models with constant boundaries (M2 and M4) provided a better fit to the observed response times compared to their counterparts with collapsing boundaries. This suggests that there might not be a substantial urgency effect under these experimental conditions.

The baseline model (M3)~\cite{mohammad2023cognitive}, which was originally developed in the context of scenarios with shorter distances and greater variability in initial velocity, effectively described 7 out of 10 qualitative patterns in our data. Model M4, in contrast, comprehensively described all qualitative patterns we observed by employing constant boundaries and including a velocity-dependent initial bias. The fitted parameters for M4 were $\alpha = 0.05$, $\beta = 0.52$, $\theta_s = 148$, $B = 1.4$, $b_z = 0.11$, $\theta_z = 8.48$, $\mu_{ND} = 0.53$, $\sigma_{ND} = 0.10$.

\section{Discussion} \label{discussion}
We conducted a driving simulator experiment to determine the effect of the oncoming vehicle type (automated vs. human-driven vehicle) and the dynamic changes in the oncoming vehicle's acceleration on human drivers' overtaking behavior. Subsequently, we used the drift-diffusion modeling framework to describe gap acceptance during these overtaking interactions.
Our experimental results revealed that gap acceptance in overtaking depends on the initial distance and TTA to the oncoming vehicle, and on the ego vehicle's initial velocity. These findings resonate with other gap acceptance studies in overtaking~\cite{farah2009passing, Farah2010, SEVENSTER2023329}. Most importantly, our study reveals two new empirical findings and one advancement in cognitive modeling: Firstly, we found no evidence of changes in participants' gap acceptance when interacting with an AV as opposed to an HDV. This finding implies that future overtaking behavior models do not necessarily need to increase their complexity by incorporating vehicle type. 
Secondly, the oncoming vehicle's acceleration profile did not affect overtaking behavior. 
Potentially, this limits the effectiveness of implicit longitudinal communication cues by AVs to show yielding behavior to human-driven vehicles during overtaking maneuvers. Finally, we showed that a version of a drift-diffusion model proposed earlier for simple overtaking scenarios can adequately describe human overtaking behavior in interactions with oncoming vehicles with varying longitudinal dynamics.

\subsection{Oncoming automated and human-driven vehicles: two peas in a pod?} \label{AV}
Studies of other gap acceptance situations showed conflicting results on whether humans change their behavior when interacting with AVs. Soni et al.~\cite{Soni2022} and Trende et al.~\cite{trende2019investigation} found that drivers were willing to accept shorter gaps at unsignalized intersections with approaching AVs, i.e., drivers decreased their critical gaps when they interacted with AVs as opposed to HDVs. However, in both studies, participants were given information about the AV's (strategic) behavior with the intention of influencing their perception of AVs. Studies that did not inform their participants about the AV's behavior showed no significant difference in gap acceptance behavior~\cite{Reddy2022, velasco2019studying, palmeiro2018interaction}. 
Similar to these studies, we did not inform participants beforehand about the AV's behavior, and found that participants did not change their overtaking behavior between sessions with an oncoming AV and an oncoming HDV. Taken together with the previous literature, this suggests that differences in human drivers' gap acceptance behavior could be attributed to the presence of bias in participants' perception of AVs; this hypothesis should be investigated in future work.
However, participants in our experiment were aware of the oncoming vehicle type (AV or HDV) before each trial due to grouping of all AV/HDV trials in a single block (which was done to streamline the experimental procedure). This prior knowledge could have restricted our ability to capture any potential bias \textit{during} the decision-making process. 
Future studies should consider randomizing the oncoming vehicle type across trials to mimic the uncertainty encountered in real mixed-traffic scenarios.

\subsection{Human overtaking behavior: insensitive to nudging?} \label{nudging}
Perhaps the most unexpected finding is that we observed no evidence for differences in participants' behavior across the acceleration profiles of the oncoming vehicle. This outcome is contrary to Rettenmaier et al.~\cite{Rettenmaier2020} who found that human drivers adapted their behavior when interacting with an oncoming vehicle with varying dynamics at a narrow passage. Consistent with this, Zgonnikov et al.~\cite{nudge} reported higher gap acceptance rates in left-turn interactions with an oncoming vehicle exhibiting a deceleration nudge profile (similar to the one considered here) compared to a constant-speed profile. The difference between these studies and our results could be potentially attributed to much larger distances investigated here (240 m to 280 m vs 50 m \cite{Rettenmaier2020} or 80 m \cite{nudge}). 

This leads us to consider two possible explanations of our findings. The first possibility focuses on human perception at the distances we examined. Despite our considered distances falling within a realistic range (e.g., \cite{llorca2016passing, Polus2000EvaluationOT}), and being comparable to those used in other simulated overtaking studies (e.g., \cite{Farah2010, leitner2023overtake, piccinini2018influence}), it may be that human drivers are inherently less sensitive to the kinds of nudges we tested. This insensitivity could be attributed to human perceptual limitations at long ranges~\cite{Schiff1990AccuracyOJ}. 
Indeed, the finding that even strong nudges (deceleration rate of $5 m/s^2$ over 2 seconds) had no significant effect on gap acceptance and response times suggests that our participants might not have perceived any change in TTA. 

Alternatively, the lack of observed behavioral differences might stem from the limitations inherent to the simulator technology used in our experiment~\cite{caird2011twelve}. While participants did perceive initial TTA discriminately, the visual resolution of our driving simulator might not have been fine enough for them to also capture subtle changes in spatial and temporal information~\cite{kemeny2003evaluating}. Future studies should explore how these simulator-specific perceptual differences may influence decision-making during overtaking maneuvers.

Besides, it is noteworthy that participants in our study fell within the age range associated with increased risky behavior in driver simulators, potentially due to videogame experience \cite{stinchcombe2017driving}. Any increased risky behavior of participants might decrease their sensitivity to the oncoming vehicle's dynamics. Investigating the relationship between videogame experience and simulator behavior can help explain individual differences in driving decisions in simulators.

\subsection{Simpler drift-diffusion models for more complex overtaking scenarios?} \label{simplermodels}
Cognitive process models offer distinct advantages over purely statistical models when analyzing human behavioral data: they provide a structured framework for comprehending the underlying cognitive mechanisms and causal relationships that drive observed behaviors. While statistical models can depict data correlations, cognitive models go a step further, allowing exploration into \textit{why} drivers accept gaps, rather than merely describing \textit{what} factors they take into account. Furthermore, cognitive models add insight into \textit{how} human drivers process relevant perceptual information over time, emphasizing the decision-making process itself.

Our study demonstrated that the cognitive modeling approach is suitable for describing human decision-making processes during overtaking interactions with oncoming vehicles with time-varying dynamics. This contributes to the existing modeling literature, which until now either considered overtaking interactions without time-varying dynamics~\cite{mohammad2023cognitive} or modeled time-varying dynamics in interactions other than overtaking~\cite{Pekkanen2022, nudge}.

Out of the four tested DDMs, only M4 was capable of describing all the qualitative patterns in our overtaking dataset. Unexpectedly, the baseline model M3 proposed by Mohammad et al.~\cite{mohammad2023cognitive}, based on a dataset involving more straightforward overtaking maneuvers~\cite{SEVENSTER2023329}, failed to describe all patterns. The key difference between M3 and M4 is that the former incorporates time-varying decision boundaries, while the latter employs more simple constant boundaries. 
Several factors could explain this difference between our work and Mohammad et al.~\cite{mohammad2023cognitive}. Firstly, even though Mohammad et al. performed model selection across 8 possible candidate models, they did not consider models with constant boundaries, which we did include here. On the other hand, Mohammad et al. reported only minor collapsing of boundaries over time ($k$ value of 0.02 in Eq.~\eqref{eq:bound_drift}), which indicated that time-varying boundaries might be redundant. To clarify this point further, we fitted the model with constant boundaries (M4) to the dataset modeled by Mohammad et al.; we found that M4 describes the dataset of~\cite{SEVENSTER2023329} better than M3, suggesting that the constant-boundary model indeed provides a more accurate and parsimonious description of overtaking gap decisions (see online supplementary information).

Secondly, the longitudinal movement profiles of the oncoming vehicle previously modeled by Mohammad et al. had constant \textit{acceleration}, as opposed to (time-varying) \textit{deceleration} modeled here; this could have led to a stronger urgency effect in the dataset they used, something that is typically associated with collapsing boundaries~\cite{ZgonnikovAbbinkMarkkula}.
Although our findings regarding the decision boundary diverge from Mohammad et al.~\cite{mohammad2023cognitive}, our results reinforced their other insights into the dynamics of the decision-making process during overtaking. Specifically, they emphasized the dynamic dependence of drift rate on both distance and time-to-arrival, as well as a velocity-dependent starting point. The fact that these conclusions persisted across studies despite differences in the experimental setup (varying TTA values, larger distances, smaller variation in the initial speed of the ego vehicle) suggests that the DDMs of the kind considered here can provide a generalizable description of human overtaking decisions and their timing.

An important limitation of our model is that it is limited to decisions and response times of only the \textit{final} decision, and in its current state cannot capture ``changes-of-mind''~\cite{resulaj_changes_2009}: the situations in which a driver initially rejected a gap but then decided to accept it after all. In our analyses, we counted these decisions as accepted gaps, disregarding participants' potential initial inclination to reject the gap. The cognitive models we analyzed here are not readily able to account for these too. Similarly, aborted overtaking maneuvers were not accounted for in our models: such aborted maneuvers represent scenarios where participants initially accepted a gap and start executing an overtaking maneuver but subsequently decided to reject it. 

Only considering final decisions restricts the predictive power of the model in regards to the decision-making \textit{process} and (to a lesser extent) the decision \textit{outcome} \cite{atiya2020changes}. For example, the decision-making process could continue even after rejecting a gap when there is late-arriving evidence in favor of accepting the gap~\cite{resulaj_changes_2009}.
Although aborted gaps have been measured before and factors (e.g., individual driver's age and gender) affecting the probability of aborting an overtaking maneuver have been studied \cite{farah2016drivers, Polus2000EvaluationOT}, this is not the case for changes-of-mind (reverting from rejecting to accepting a gap). Therefore understanding and modeling these types of overtaking decisions requires further research, including extensions of models like DDM and corresponding fitting tools to incorporate changed decisions.

\subsection{Closing the gap: implications of cognitive modeling of intricate traffic decisions}
Understanding and modeling human behavior in dynamic interactions between AVs and human drivers is essential for safe transportation systems of the future~\cite{Schieben2019}. For instance, AVs can improve their own decision-making and planning by incorporating predictions of human road user behavior~\cite{schumann2023using}. Here, an AV can adopt the perspective of human-driven vehicles and employ perceptual cues such as distance and TTA in the simulated human drivers' evidence accumulation process to predict the likelihood and timing of their gap acceptance, to adjust its behavior accordingly. Yet, determining the exact initiation point of the decision-making process remains a complex task, as we currently assume that the desire to perform a particular maneuver already exists. 
 
Finally, cognitive models like DDMs can contribute to more realistic simulations of human-AV interactions \cite{guido2019using, markkula2018models}. These models can be embedded in the trajectory control of human-driven vehicles in microscopic traffic simulations, allowing for rigorous training and testing of AV performance within highly realistic simulated environments. This becomes particularly valuable when training and validation data are scarce or when certain scenarios are deemed too dangerous for data collection in real-world interactions with AVs, which can often be the case for overtaking \cite{Trafton2020}. However, a major challenge in portraying realistic scenarios is simulating the impact of individual differences and how parameters of the cognitive model change over time, potentially influenced by humans' perception of AVs in emerging mixed traffic. In our study, participants on average did not change their overtaking behavior when interacting with an AV (as compared to interactions with an HDV), although we did find evidence of individual differences (see online supplementary information). However, as human drivers become increasingly exposed to human-AV interactions on the road in the future, new behavioral patterns may evolve. Therefore, continued empirical and modeling research is essential to ultimately unlock the full potential of cognitive process models for automated vehicle development.

\addtolength{\textheight}{-1cm}
\section*{CRediT authorship contribution statement}
\textbf{Samir H.A. Mohammad:} Conceptualization, Data curation, Formal analysis, Investigation, Methodology, Project administration, Software, Validation, Visualization, Writing – original draft, Writing – review \& editing. \textbf{Haneen Farah:} Conceptualization, Methodology, Resources, Supervision,
Writing – review \& editing. \textbf{Arkady Zgonnikov:} Conceptualization, Methodology, Project administration, Resources, Supervision, Writing – original draft, Writing – review \& editing.

\section*{Declaration of competing interests}
The authors declare that they have no known competing financial interests or personal relationships that could have appeared to influence the work reported in this paper.

\section*{Data availability}
All the data and code produced in this study, as well as online supplementary information are available at \href{https://osf.io/p2wme}{https://osf.io/p2wme}.

\section*{Acknowledgments}
We would like to thank Olger Siebinga for providing technical support with JOAN and Federico Scarì and Riccardo Sepe for their help with experiment design.

%%%%%%%%%%%%%%%%%%%%%%%%%%%%%%%%%%%%%%%%%%%%%%%%%%%%%%%%%%%%%%%%%%%%%%%%%%%%%%%%

%\begin{thebibliography}{99}
% \printbibliography

\bibliographystyle{jabbrv_ieeetr}
\bibliography{IEEEabrv,report}
%\end{thebibliography}
\end{document}